\documentclass[%
 reprint,
showpacs,preprintnumbers,
 amsmath,amssymb,
 aps,
pre,
]{revtex4-1}

\usepackage{graphicx}
\usepackage{dcolumn}
\usepackage{bm}

\begin{document}

\preprint{APS/123-QED}

\title{Boussinesq approximation of the Cahn-Hilliard-Navier-Stokes equations}

\author{Anatoliy Vorobev}
 \email{A.Vorobev@soton.ac.uk}
\affiliation{Energy Technology Research Group, School of Engineering Sciences, \\ University of Southampton, SO17 1BJ, UK}

\date{\today}

\begin{abstract}
We study the interactions between the thermodynamic transition and hydrodynamic flows which would characterise a thermo- and hydro-dynamic evolution of a binary mixture in a dissolution/nucleation process. The primary attention is given to the slow dissolution dynamics. The Cahn-Hilliard approach is used to model the behaviour of evolving and diffusing interfaces. An important peculiarity of the full Cahn-Hilliard-Navier-Stokes equations is the use of the full continuity equation required even for a binary mixture of incompressible liquids, firstly, due to dependence of mixture density on concentration and, secondly, due to strong concentration gradients at liquids' interfaces. Using the multiple-scale method we separate the physical processes occurring on different time scales and, ultimately, provide a strict derivation of the Boussinesq approximation for the Cahn-Hilliard-Navier-Stokes equations. This approximation forms a universal theoretical model that can be further employed for a thermo/hydro-dynamic analysis of the multiphase systems with strongly evolving and diffusing interfacial boundaries, i.e. for the processes involving dissolution/nucleation, evaporation/condensation, solidification/melting, polymerisation, etc. 
\end{abstract}

\maketitle


\section{Introduction}

Fluid mixtures can be classified as being immiscible (e.g. oil/water mixture) or miscible (e.g. honey/water mixture or mixture of two gases). An immiscible system is characterised by a strict interfacial boundary. This boundary cannot be crossed by the molecules of adjoining liquids, which, on molecular scale, can be explained by a high potential barrier due to different intermolecular interactions within the mixture components. At macro-scale, the coefficient of surface tension is introduced to define the macroscopic effects of the molecular potential barrier.

The mixture of two gases is an example of a directly opposite case. As intermolecular forces between gas molecules are negligibly small, no potential barrier at the gases' boundary exists, and gas molecules of initially separated components freely co-diffuse. So there is no sense in introduction of an interface, and hence, there is no surface tension on the gases' boundary.

The focus of the current work is on the miscible mixtures of two liquids, for which the intermolecular forces cannot be neglected. As an everyday example, the behaviour of a droplet of honey in water may be considered: for such a droplet, a strict interface is visible for a long period after immersion of a droplet into water; but after a slow dissolution process honey/water mixture becomes homogeneous. Again, the existence of a strict interface between mixture components, on molecular scale, should be associated with a potential barrier. But, in comparison with the immiscible case, for miscible interfaces, some molecules with sufficiently high kinetic energies are able to cross this barrier. It may be further assumed that the molecular-flux through an interface gradually diminishes the barrier's height which would result in growing numbers of molecules being diffused from one phase to another, and, ultimately, in a complete dissolution of a droplet.

Thus, the concept of interface is required to describe the behaviour of a slowly miscible system. The interface, at macro-scale, will be characterised by the surface tension coefficient. In contrast to immiscible systems, surface tension coefficient varies as the dissolution process progresses, see e.g. Ref. \onlinecite{Joseph93}. The cases of completely and partially miscible liquids (e.g. honey/water and butanol/water mixtures) would differ from each other by the fact that the surface tension coefficient decreases to zero in the first case (as the interface disappears) and is dynamically variable over dissolution process but always remains non-zero in the second case.

Concluding this discussion, we would like to state that the mixing of two (even completely miscible) liquids cannot be reduced to a simple diffusion (as it can be done for mixing of two gases). The surface tension effects have a two-fold influence on the dissolution dynamics, both the morphology of the interface and the rate of mass transfer through the interface are affected. We may argue that the surface tension is sufficiently high to exclude the co-diffusion of molecules between immiscible liquids. In the case of miscible systems, the mass transfer through the interfacial boundary is not zero but its rate is restricted by the surface tension effects. 

A first review devoted to the physics of slowly miscible systems can be found in Ref. \onlinecite{Joseph93}. The phase-field approach was adopted to model a hydrodynamic evolution of a miscible interface. The authors also pointed out another physical effect that should characterise evolution of binary mixtures, namely, the quasi-compressibility of the hydrodynamic equations: even for a mixture of two incompressible liquids, the fluid velocity is a non-solenoidal field due to dependence of mixture density on concentration.  

The first comprehensive numerical studies for miscible interfaces were however based on the governing equations in which the transport of a second component was modelled as an impurity \cite[][]{Chen96}, i.e. with no account for the surface tension effects. The Cahn-Hilliard-Navier-Stokes equations, that include all essential physical effects previously discussed by Joseph and Renardy, have been derived in Ref. \onlinecite{Lowengrub98}. These equations fully define the hydrodynamic behaviour of miscible binary mixtures.  Nevertheless, the recent numerical studies of the miscible multiphase systems are still based on the impurity-like equations, in which however the Korteweg stress has been added into the momentum balance \cite[][]{Chen02, Bessonov04, Chen07}. We should note that the Korteweg stress is only one of the surface tension effects, since the rate of phase change is also determined by the surface tension effects, which however is not taken into account in any recent hydrodynamic model for a miscible multiphase system. 

Hydrodynamics of immiscible systems has been studied either for density-matched fluids \cite[][]{Chella96, Badalassi03, Kim2004}, for such fluids, the quasi-compressibility effects are eliminated, or by using the Boussinesq-like approximation \cite[][]{Jacqmin99, Jacqmin00, Villanueva06, Ding07a, Ding10}. Since within the phase-field approach it is necessary to assume that the density gradients are large at least in some parts of a computational domain, this makes justification of the Boussinesq approximation difficult. The role of the quasi-compressibility effects for a particular system was examined in Refs. \onlinecite{Joseph96} (for an analysis of one-dimensional diffusion within a pipe) and  \onlinecite{Chen02} (for an analysis of miscible displacements in capillary tubes). Following Ref. \onlinecite{Joseph96}, the velocity field was divided into incompressible and expansion parts. In both works, the estimations showed that the expansion part of the velocity field is negligibly small. Nevertheless, even such a statement is not sufficient to prove the use of the Boussinesq approximation. At present, we are unaware of any paper where the Boussinesq approximation of the full Cahn-Hilliard-Navier-Stokes equations has been strictly derived. 

To conclude this introductory section, we would like to mention that applications that involve the miscible interfaces are ubiquitous. They include solvent extraction, cleaning, removal of oil spills, waste treatment, enhanced oil recovery, drug delivery, etc. Frequently, it is assumed that the rate of dissolution and the rate of flow change are not comparable, e.g. dissolution is a slow process and the changes, caused by hydrodynamic flows, happen at much faster rate, or, on the contrary, nucleation is a very fast process and the hydrodynamic flows are much slower. For such cases, different simplified models were proposed which allow these processes to be considered separately. We however are interested in slow flows when the typical dissolution and convective time-scales are comparable; an example can serve the hydrodynamic flows in capillary tubes or in porous media. Other important examples where the thermodynamic and hydrodynamic processes interact are the flows in near-critical systems.

\section{Cahn-Hilliard-Navier-Stokes-Equations}
\subsection{Phase-field approach}

There are several different approaches used to describe the behaviour of a multiphase system. An immiscible interface has usually very small thickness (of several molecular layers), which makes the use of the classical Laplace approach to model the interface as a surface of discontinuity (of zero thickness) being well justified. Following this approach, two systems of Navier-Stokes equations are separately solved in each phase and so obtained solutions are matched using boundary conditions. This approach is however difficult to apply when interfacial boundaries undergo complex (topological) modifications. For similar problems, the phase-field (also called the diffuse-interface) approach becomes more suitable. 

The main idea of the phase-field approach is to artificially smear the interfacial boundary and one system of Navier-Stokes equations is then solved for the entire multiphase system. All variables experience strong but continuous changes through the interface. The position and shape of the interface is tracked either by introduction of additional scalar function (e.g. level-set function in the level-set method) or by using a natural variable (like concentration field). A new volume force is also introduced to mimic the surface tension effects. The fluid behaviour in the limit of zero surface thickness is usually analysed. We however should notice that the interface smearing is physically justified for the systems near thermodynamic critical point.

In the current work, the phase-field approach is utilised to define the evolution of a multiphase system. The mixture components are called solvent and solute. To characterise a state of the binary mixture the concentration field is used, which we define as the mass fraction of the solute in the solvent phase.

\subsection{Thermodynamic model}

To define a thermodynamic state of a binary mixture we need to introduce the free energy function. This function can be either written based on experimental data or derived from a molecular level theory. Only for illustration purposes, let us show how the free energy of a heterogeneous system should be defined if the classical Laplace approach is used. In this approach the phases are treated separately and the free energy function includes three terms,
\begin{eqnarray}
F=V_{1}f_{1}(\rho_{01})+V_{2}f_{2}(\rho_{02})+S(V_{1}),\\ S(V_{1})\equiv\sigma(36\pi)^{1/3}V_{1}^{2/3}.
\end{eqnarray}
Here, the first and second terms are, respectively, the free energies of the first and second components of a mixture. The last term, $S(V_{1})$, is an interfacial energy traditionally defined through the surface tension coefficient $\sigma$. $V_{1}$ and $V_{2}$ are the volumes of the first and second phases.

In the phase-field approach, one system of equations is used to define the behaviour of an entire multiphase system. It is convenient to introduce the specific free energy function $f$. To take the surface tension effects into account, Cahn and Hilliard \cite[][]{Cahn58} proposed to define $f$ not only as a function of density and concentration but also as a function of concentration gradient,
\begin{equation}\label{free_energy}
  f(\rho,C,\nabla C)=f_0(\rho,C)+\frac{\epsilon}{2} | \nabla C | ^2.
\end{equation}
Here, $f_0$ is called the classical part of the free energy, $C$ is the solute concentration, and $\epsilon$ is a constant called the capillary coefficient. Coefficient $\epsilon$ is assumed to be very small, so the second term is not-negligible only at the places of strong gradients of concentration, i.e. at interfaces.

The chemical potential $\mu$, derived from the free energy function (\ref{free_energy}) with the use of assumption of incompressibility of mixture components (i.e. $\rho$ is function of concentration $C$ and independent of pressure $p$) reads \cite[][]{Lowengrub98} 
\begin{eqnarray}\label{chem_pot}
  \mu(p,C) = \mu_0(C) - \frac{p}{\rho^2} \frac{d \rho}{d C} - \frac{\epsilon}{\rho} \nabla\cdot (\rho \nabla C),\\ \mu_0(C)\equiv\frac{df_0(C)}{dC}.
\end{eqnarray}
Here $\mu_0$ stands for the classical part; the incompressibility assumption brings an explicit dependence of the chemical potential on pressure; and the last term is a non-classical contribution which stems from the Cahn-Hilliard addition in (\ref{free_energy}).

\begin{figure}
  \centerline{\includegraphics{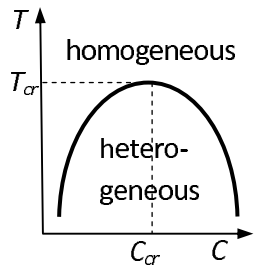}}
  \caption{The phase diagram for a binary system with an upper critical solution temperature.}\label{fig:phase_diagram}
\end{figure}

The expression to be adopted for the classical part of the free energy function, $f_0$, is written so as to reproduce the behaviour defined by the phase diagram depicted in figure \ref{fig:phase_diagram}. In general, there are other different types of phase diagrams characterising the states of binary systems \cite[][]{LandauV}. The diagram depicted in figure \ref{fig:phase_diagram} is however one of the most popular. This diagram defines the mixture with the upper critical solution temperature: a system with the temperature below the critical point may be either homogeneous or heterogeneous depending on concentration (the amount of a second component), while a supercritical system is always homogeneous.

In the current work, we use the expression for the free energy function originally proposed by Landau \cite[][]{LandauV} to define a thermodynamic state of a system near its critical point,
\begin{equation}\label{f_0}
  f_{0}(C) = a(C-C_{cr})^2+b(C-C_{cr})^4. 
\end{equation}
In this expression, $C_{cr}$ is the solute concentration in the critical point, and coefficients $a$ and $b$ are the phenomenological parameters which define the choice of a particular binary mixture. It is necessary to note that the coefficient $a$ absorbs the factor $(T-T_{cr})$, i.e. (i) $a$ tends to zero as the system approaches the critical point, and (ii) $a$ is negative for undercritical conditions and positive for the temperatures above the critical value. Function (\ref{f_0}) has two minima for negative values of $a$ and one minimum for positive $a$. An equilibrium of a thermodynamic system corresponds to a minimum of the free energy function. Two minima characterising a system below the critical point are associated with two different phases, while in supercritical conditions a binary mixture is homogeneous and is characterised by the only minimum of the free energy function.   

Using function (\ref{f_0}) we may derive an expression for the classical part of the chemical potential,
\begin{equation}
 \mu_0(C)=2a(C-C_{cr})+4b(C-C_{cr})^3.
\end{equation}
If, firstly, the surface tension is not taken into account (the Maxwell theory is adopted) and, secondly, there is no loss or gain in volume upon liquids' mixing (simple mixtures are considered), then the equilibrium concentrations of the mixture components are defined by $C_{01,02}=C_{cr}\pm(-\frac{a}{2b})^{1/2}$.

We will use expression (\ref{f_0}) as a simple model for a binary system that, under different conditions, may define both miscible and immiscible solutions. Another popular choice for the free energy function frequently used to model the behaviour of immiscible mixtures reads
\begin{equation}\label{Jacq_f0}
  f_0 = \frac{1}{2}(C-C_{01})^2(C-C_{02})^2. 
\end{equation}
It can be shown that (\ref{Jacq_f0}) transforms into (\ref{f_0}) if $b=1/2$ and $C_{01,02}\equiv C_{cr}\pm(-a)^{1/2}$. That is, formula (\ref{f_0}) can be used to define the binary mixtures even under temperatures far from the critical one, where it can be considered as an interpolation of the experimental data; two new phenomenological coefficients $a$ and $b$ are determined so as to provide the best fit to the experimental data.

Finally, we notice that a thermodynamic model defines the equilibrium states. A thermodynamic system can reach its equilibrium following a long equilibration process (complete dissolution takes hours for the honey/water mixture). In the present work we aim to define the behaviour of a binary system during equilibration to its thermodynamically stable state.

\subsection{Hydrodynamic model}

The evolution of a binary mixture to its thermodynamic equilibrium is defined by the hydrodynamic model. The governing equations for hydrodynamic evolution of the Cahn-Hilliard fluid were derived in Ref. \onlinecite{Lowengrub98}. These equations include the laws of conservation of mass, species and momentum,  
\begin{equation}\label{continuity}
  \frac{\partial \rho}{\partial t} + \nabla\cdot (\rho \boldsymbol{v})=0, 
\end{equation}
\begin{equation}\label{species_cons}
  \rho \left( \frac{\partial C}{\partial t}+(\boldsymbol{v}\cdot\nabla) C \right) = \alpha \nabla^2 \mu,
\end{equation}
\begin{equation}\label{CHNS}
  \rho\left(\frac{\partial \boldsymbol{v}}{\partial t} +(\boldsymbol{v}\cdot\nabla) \boldsymbol{v}\right)=- \nabla p + \nabla \cdot \tau_{\eta} - \epsilon \nabla \cdot \tau_{\epsilon}.
\end{equation}
Here, the constant $\alpha$ is the coefficient of mobility. 

The fluids studied in this paper are assumed to be incompressible, but due to large concentration gradients at the interface the continuity equation (\ref{continuity}) needs to be taken in its full form. Such fluids are called quasi-incompressible \cite[][]{Joseph93, Lowengrub98}. The dependence of the fluid density $\rho$ on the solute concentration $C$ has the simplest form for the so-called simple mixtures, for which
\begin{equation}\label{density}
  \frac{1}{\rho}=\frac{1-C}{\rho_{01}}+\frac{C}{\rho_{02}},
\end{equation}
where $\rho_{01}$ and $\rho_{02}$ are the solvent and solute densities. So for the mixture of water and glycerol, expression (\ref{density}) is true with 1\% accuracy; for the mixture of methanol and water, the accuracy of (\ref{density}) is 3.5\% \cite[][]{Joseph93}. 

As the velocity field is non-solenoidal, the divergence term in the expression for the viscous stress tensor $\tau_\eta$ also needs to be taken into account,
\begin{equation}\label{visc_stress}
  \tau_\eta=\eta\left(\nabla\otimes\boldsymbol{v}+(\nabla\otimes\boldsymbol{v})^T-\frac{2}{3} (\nabla\cdot\boldsymbol{v}) I \right).
\end{equation}
In this expression, the symbol $\otimes$ stands for the tensor product, the index $T$ denotes the transposed, and $I$ denotes the unit tensor. The viscosity coefficient $\eta$, introduced by expression (\ref{visc_stress}), should, in general, depend on the solute concentration, and can be represented by a simple mass-weighted approximation or by an experimental formula. 

The tensor $\tau_\epsilon$, defined as
\begin{equation}
  \tau_\epsilon=\rho\nabla C \otimes \nabla C,
\end{equation}
introduces the Korteweg interfacial stress. 

The right hand side of equation (\ref{species_cons}) takes into account the species transport due to diffusion. Here, the driving force for the diffusion flux is the gradient of chemical potential. The classical Fick's law which states a linear proportionality between a diffusive flux and concentration gradient is only valid for weak solutions with small concentration gradients and cannot be used in our case \cite[][]{LandauV}. One should also notice, that diffusive term of equation (\ref{species_cons}) includes the pressure and surface tension effects (stemming from expression (\ref{chem_pot}) for the chemical potential).
 
The full system of equations, defined in this section, is complemented with boundary and initial conditions. As initial conditions, the velocity $\boldsymbol{v}$ and concentration $C$ are assumed to be known. As boundary conditions for the velocity field, the standard no-slip condition is used. For the field of solute concentration, two conditions need to be imposed at each boundary as (\ref{chem_pot}) and (\ref{species_cons}) define the fourth order system of equations in terms of concentration field $C$. The first boundary condition is the zero-flux requirement,
\begin{equation}
  \boldsymbol{n}\cdot\nabla \mu =0.
\end{equation}
Here $\boldsymbol{n}$ is the vector normal to the boundary. The second boundary condition imposes the local thermodynamic equilibrium of the fluid-wall system. The full form of this condition is discussed in the works of \cite{Jacqmin00} and \cite{Ding07b}. For the simplest case of the contact angle of $90^0$, this condition can be written as
\begin{equation}\label{bc_90}
  \boldsymbol{n}\cdot\nabla C =0.
\end{equation}
In such a form, this condition states that the interfacial boundary is orthogonal to the wall, i.e. molecules of the wall are neutral to components of a binary mixture.

At the end of this section, let us enumerate the essential features required for derivation of governing equations (\ref{continuity}), (\ref{species_cons}) and (\ref{CHNS}). The first two requirements have been already mentioned, they are the use of the Cahn-Hilliard free energy function (\ref{free_energy}) and the assumption that the mixture components are incompressible liquids. The third requirement is to define the velocity, $\boldsymbol{v}$, as the mass-averaged quantity over random velocities of different molecules that constitute a fluid particle. The volume-averaged definition for the fluid velocity of a multiphase system is also possible and can even have some advantages as it would automatically produce a divergence-free velocity field for the mixing of two simple incompressible liquids \cite[][]{Boyer02, Ding07a}. However, a traditional definition for the velocity in fluid mechanics is the mass-averaged quantity. The resultant governing equations based on this definition can be more easily and more naturally compared with the classical equations for a single phase medium. 

\section{Separation of time-scales}

The quasi-compressibility of equations (\ref{continuity}), (\ref{species_cons}) and (\ref{CHNS}) brings short-term processes. This significantly complicates the numerical solution of equations. The slow diffusion and convective evolution of a binary system would frequently be a primary practical interest, and even if we are not interested in quick processes, the numerical integration with a time step smaller than the smallest typical time scale is required in order to obtain a convergent numerical solution. In this section, the time-scales that characterise the evolution of a multiphase system are first identified and then, by using the multiple-scale method, the general governing equations are split into the systems which separately define the physical processes on different time-scales.

\subsection{Non-dimensionalization of governing equations}

For further analysis we re-define the density and concentration fields by shifting their reference points by the critical values, $\rho_{cr}$ and $C_{cr}$, namely,
\begin{equation}
  (\rho - \rho_{cr}) \rightarrow \rho,\quad (C-C_{cr}) \rightarrow C.
\end{equation}
We also non-dimensionalize the governing equations using the following units of time $\tau_*$, velocity $V_*$, pressure $p_*$ and specific free energy:
\begin{equation}
  \tau_*=\frac{L_*}{V_*},\quad V_*=\mu_*^{1/2}, \quad p_*=\rho_*\mu_*,\quad f_*=\mu_*.
\end{equation}
Here $L_*$ is the typical size, $\rho_*$ is the typical density (e.g. $\rho_{cr}$), and $\mu_*$ is the unit of the chemical potential, for which we can use $\mu_*=b$. The resultant dimensionless equations then read
\begin{equation}\label{cont_nondim}
  \frac{\partial \rho}{\partial t} + (\boldsymbol{v}\cdot \nabla)\rho = -(1+\rho)\nabla\cdot\boldsymbol{v},
\end{equation}
\begin{equation}\label{poisson}
  (1+\rho)\left(\frac{\partial C}{\partial t} + (\boldsymbol{v}\cdot \nabla) C\right) = \frac{1}{Pe}\nabla^2\mu,
\end{equation}
\begin{eqnarray}\label{NS_nondim}
  (1+\rho) \left( \frac{\partial \boldsymbol{v}}{\partial t} + (\boldsymbol{v} \cdot \nabla) \boldsymbol{v} \right) = \nonumber \\ 
   -\nabla p + \frac{1}{Re} \nabla \cdot \tau_\eta - Ca\nabla \cdot \tau_\epsilon - Ga (1+\rho)\boldsymbol{\gamma}, 
\end{eqnarray} 
\begin{equation}\label{pressure}
  p = \frac{1}{\phi} \left( -\mu + \mu_0 - Ca \left[\triangle C + \phi(1+\rho)(\nabla C)^2\right] \right),
\end{equation}
\begin{equation}
  f_0=AC^2+BC^4,\quad \mu_0=\frac{d f_0}{dC},
\end{equation}
\begin{equation}
  \rho=\frac{\phi C}{1-\phi C},  
\end{equation}
\begin{equation}\label{tau_eta_nondim}
  \tau_\eta = \eta\left(\nabla\otimes\boldsymbol{v}+(\nabla\otimes\boldsymbol{v})^T-\frac{2}{3} (\nabla\cdot\boldsymbol{v}) I\right),
\end{equation}
\begin{equation}\label{Korteweg_nondim}
  \tau_\epsilon = (1+\rho)\nabla C \otimes \nabla C.
\end{equation}
The last term in equation (\ref{NS_nondim}) is the gravitational force (with $\boldsymbol{\gamma}$ being a unit vector directed upwards). A non-dimensionalised coefficient of viscosity $\eta$ (in formula (\ref{tau_eta_nondim})) is obtained using the typical value $\eta_*$ (e.g. the viscosity coefficient of a binary mixture in its critical point).   

The non-dimensional parameters entering the above equations are the Peclet number
\begin{equation}
  Pe=\frac{\rho_*L_*}{\alpha\mu_*^{1/2}},
\end{equation}
the capillarity parameter
\begin{equation}
  Ca=\frac{\epsilon}{\mu_*L_*^2},
\end{equation}
the Galileo number
\begin{equation}
  Ga=\frac{g L_*}{\mu_*},
\end{equation}
and the Reynolds number
\begin{equation}
  Re=\frac{\rho_*\mu_*^{1/2}L_*}{\eta_*}.
\end{equation}

The binary mixture is also characterised by the parameter
\begin{equation}\label{phi}
  \phi=\frac{\rho_*}{\rho_{01}}-\frac{\rho_*}{\rho_{02}},
\end{equation}
where $\rho_{01}$ and $\rho_{02}$ are the densities of the mixture components. The adopted expression for the free energy function also contains another non-dimensional parameter that defines a distance from the critical point,
\begin{equation}\label{A}
  A=\frac{a}{\mu_*}.
\end{equation}
This parameter also defines whether the system is heterogeneous (negative A) or homogeneous (positive A) in equilibrium.

Finally, let us also write the expression for the total energy of the fluid enclosed by volume $V$,
\begin{equation}
  E=\int_V (1+\rho)\left[\frac{\boldsymbol{v}^2}{2}+f_0+\frac{Ca}{2}(\nabla C)^2+Ga (\gamma\cdot\boldsymbol{r})\right]dV.
\end{equation}
Here the first term corresponds to the fluid kinetic energy, the second term is the classical part of free energy, the third term accounts for the energy accumulated by interfaces and the last term is the potential energy due to gravity. The total energy is measured in the units of $\rho_* \mu_* L_*^3$.

\subsection{Typical time scales}
To identify the different time scales characterising the physical system defined by the full equations written in the previous section we consider an evolution of a one-dimensional (along the $x$-axis) small-amplitude perturbation on the background of a stationary homogeneous state. For simplicity, we also omit the gravity term for this analysis. 

Evolution of such a perturbation is defined by the following linearised equations
\begin{eqnarray}
\frac{\partial \rho}{\partial t}=-\frac{\partial v}{ \partial x}, \\
\frac{\partial C}{\partial t}=\frac{1}{Pe}\frac{\partial^2 \mu}{\partial x^2}, \\
\frac{\partial v}{\partial t}=-\frac{\partial p}{\partial x}+\frac{4}{3 Re}\frac{\partial^2 v}{\partial x^2}, \\
p=\frac{1}{\phi}\left(-\mu+2AC-Ca\frac{\partial^2 C}{\partial x^2}\right),\quad \rho=\phi C.
\end{eqnarray}
Seeking a solution in the form of a plane wave, $C \sim \exp(ikx-i\omega t)$, where $i$ is the imaginary unit, $k$ is the wave-length and $\omega$ is the frequency, we obtain the following dispersion relation:
\begin{equation}
\omega^2=-i\omega\left(\frac{Pe}{\phi^2}+\frac{4}{3Re}k^2\right)+\frac{k^2}{\phi^2}(2A+Ca k^2).
\end{equation} 
Next, we split $\omega$ into the real and imaginary parts, $\omega=\omega_r+i\omega_i$ and equate the coefficients in front of the like terms to obtain 
\begin{eqnarray}\label{full_disp}
(\omega_r^2-\omega_i^2)=\nonumber \\ \omega_i\left(\frac{Pe}{\phi^2}+\frac{4}{3Re}k^2\right)+ \frac{k^2}{\phi^2}(2A+Cak^2), \\
2\omega_r\omega_i=-\omega_r\left(\frac{Pe}{\phi^2}+\frac{4}{3Re}k^2\right).
\end{eqnarray}
The analysis of the second equation shows that this is satisfied by either 
\begin{equation}\label{1sol}
\omega_r=0,
\end{equation}
or
\begin{equation}\label{2sol}
\omega_i=-\frac{1}{2}\left(\frac{Pe}{\phi^2}+\frac{4}{3Re}k^2\right).
\end{equation}

To simplify further derivations we will use that
\begin{equation}\label{inequality}
\frac{k}{\phi}(2A+Ca k^2)^{1/2} \ll \left(\frac{Pe}{\phi^2}+\frac{4}{3Re}k^2\right);
\end{equation}
or, equivalently,
\begin{equation}\label{inequalities}
\phi \ll 1,\quad A\ll 1,\quad Ca \ll 1,\quad Pe \sim 1.  
\end{equation}
That is, we assume a small density difference between components of binary mixtures, closeness to the critical point and the existence of an interface. 

Substitution of condition (\ref{1sol}) into (\ref{full_disp}) produces two formulae for the decrements which define the monotonic decay of the considered 1D perturbation, 
\begin{eqnarray}\label{omega1}
\omega_{i,1}=-\left(\frac{Pe}{\phi^2}+\frac{4}{3Re}k^2\right),\\ \label{omega2}
\omega_{i,2}=-\frac{\frac{k^2}{\phi^2}(2A+Ca k^2)}{\frac{Pe}{\phi^2}+\frac{4}{3Re}k^2}. 
\end{eqnarray}

Next, we should note that equation (\ref{full_disp}) has no solution if option (\ref{2sol}) to satisfy the imaginary part of the dispersion relation is chosen.

Derived expressions (\ref{omega1})-(\ref{omega2}) allow us to conclude that the evolution of a multiphase system defined by equations (\ref{cont_nondim}), (\ref{poisson}) and (\ref{NS_nondim}) is characterised by three different time-scales, 
\begin{equation}\label{time_scales}
\tau_d=\frac{Pe}{k^2(2A+Cak^2)},\quad\tau_c=\frac{3Re}{4k^2},\quad \tau_{e}=\frac{\phi^2}{Pe}.
\end{equation} 
Here, $\tau_d$ and $\tau_c$ are the diffusion and convection time-scales, and $\tau_{e}$ is the fast time scale that we call the expansion time scale.  

As it is clear now, the inequality (\ref{inequality}) reflects the ratio between the quick and slow time-scales. For further analysis we will introduce the small parameter $\chi$ defined as
\begin{equation}
\chi^4\equiv\frac{\tau_{e}}{\tau_d}=\frac{\phi^2 k^2(2A+Ca k^2)}{Pe^2}.
\end{equation}
We are interested in dissolution dynamics, i.e. in evolution of the system on the slow diffusive time-scale.  

The main idea of the multiple time-scale method is to represent the time derivative as
\begin{equation}
\frac{\partial}{\partial t} = \frac{1}{\chi^2}\frac{\partial}{\partial t_{-2}} + \chi^2 \frac{\partial}{\partial t_2},
\end{equation}   
i.e. to explicitly introduce two times, the quick one, $t_{-2}$, describing evolution of a system on the fast expansion time-scale and the slow one, $t_2$, defining the diffusive and convective evolution. 

We will also expand all variables in the series of small parameter $\chi$,
\begin{eqnarray}
\boldsymbol{v}=\chi^2\boldsymbol{v}_2+\chi^4\boldsymbol{v}_4+\dots,\\ 
p=p_0+\chi^2p_2+\chi^4 p_4+\dots, \\
\mu=\chi\mu_1+\chi^3\mu_3+\dots, \\
\rho=\chi^2\rho_2+\chi^4\rho_4+\dots,\\ 
C=\chi C_1+\chi^3C_3+\dots,\\
\eta=1+\dots;
\end{eqnarray}
and relate the non-dimensional parameters to different orders of $\chi$,
\begin{eqnarray}\label{ratios_conv}
\phi=\phi_1\chi,\quad A=A_2\chi^2,\quad Ca=Ca_2\chi^2,\\
Re=Re_{-2}\frac{1}{\chi^2},\quad Pe=Pe_0,\quad Ga=Ga_2\chi^2.
\end{eqnarray}
We should note that there are two main conditions implying the choices for the above-written ratios, namely, (i) to save all essential physical effects and (ii) to have a closed system of governing equations. The non-dimensional parameters define the ratios between different terms in the governing equations. For derivations, we always assume that all effects in the original equations (\ref{cont_nondim}), (\ref{poisson}) and (\ref{NS_nondim}) may be important and such values of the non-dimensional parameters are chosen which permit us to include all corresponding terms in the final equations. Ratios (\ref{ratios_conv}) imply that $\tau_d$ and $\tau_c$ are of the same order. This only means that the final equations will include both convective and diffusive terms. Following such an approach, we will obtain a comprehensive theoretical model that can be used for the analysis of a wide range of problems. For a particular problem, some terms in the model may be found either small or very large, which would, sometimes, lead to further simplifications. But our aim is to obtain a general model.

\subsection{Separation of the processes occuring on different time scales}
First, we write down the different orders of equations (\ref{cont_nondim}), (\ref{poisson}) and (\ref{NS_nondim}). The first orders of the mass (\ref{cont_nondim}) and species balances (\ref{poisson}) are as follows
\begin{eqnarray}
\frac{\partial \rho_2}{\partial t_{-2}}=0,\quad\frac{\partial C_1}{\partial t_{-2}}=0; \label{rho2C1}\\
\frac{\partial \rho_4}{\partial t_{-2}}=-\nabla\cdot\boldsymbol{v}_2,\quad \frac{\partial C_3}{\partial t_{-2}}=\frac{1}{Pe_0}\nabla^2\mu_1; \\
\frac{\partial C_5}{\partial t_{-2}}+\frac{\partial C_1}{\partial t_2}+(\boldsymbol{v}_2\cdot\nabla)C_1+\rho_2\frac{\partial C_3}{\partial t_{-2}}= \nonumber \\  \frac{1}{Pe_0}\nabla^2\mu_3.
\end{eqnarray}
The first orders of the equation of momentum balance (\ref{NS_nondim}) read
\begin{eqnarray}
\frac{\partial \boldsymbol{v}_2}{\partial t_{-2}}=-\nabla p_0, \\
\frac{\partial \boldsymbol{v}_4}{\partial t_{-2}}=-\nabla p_2-Ga_2 \boldsymbol{\gamma}, \\
\frac{\partial \boldsymbol{v}_6}{\partial t_{-2}}+\frac{\partial \boldsymbol{v}_2}{\partial t_2} + (\boldsymbol{v}_2\cdot\nabla)\boldsymbol{v}_2 + \rho_2\frac{\partial \boldsymbol{v}_4}{\partial t_{-2}}= \nonumber \\-\nabla p_4+\frac{1}{Re_{-2}}\tau_{\eta,2}-Ca_2\tau_{\epsilon,2}-Ga_2\rho_2\boldsymbol{\gamma}.
\end{eqnarray}
Here, the corresponding orders of the viscous stress tensor and of the Korteweg tensor are
\begin{eqnarray}
\tau_{\eta,2}=\nabla\otimes\boldsymbol{v}_2+(\nabla\otimes\boldsymbol{v}_2)^T-\frac{2}{3}(\nabla\cdot\boldsymbol{v}_2)I,\\
\tau_{\epsilon,2}=\nabla C_1\otimes\nabla C_1.
\end{eqnarray}

In these equations, all variables are assumed to be functions of both times, $t_{-2}$ and $t_2$. Next, we will split out the processes on different time-scales using the averaging procedure briefly outlined in the next two paragraphs. 

Firstly, we assume that all fields can be split into slowly- and quickly-changing parts,
\begin{eqnarray}
\boldsymbol{v}=\boldsymbol{u}(t_{2})+\boldsymbol{w}(t_{-2},t_2),\\
\rho=\bar{\rho}(t_{2})+\tilde{\rho}(t_{-2},t_2). \label{rho_averaging}
\end{eqnarray}
Here, $\boldsymbol{u}\equiv\frac{1}{T_{2}}\int_0^{T_{2}}{\boldsymbol{v}dt}$ and $\boldsymbol{w}\equiv\boldsymbol{v}-\boldsymbol{u}$. That is, $\boldsymbol{u}$ is the fluid velocity averaged over long time-scale ($T_2$ is a time interval much larger than the fast time scale) and $\boldsymbol{w}$ defines the quick-time-scale fluctuations of the fluid velocity. For scalar quantities, the barred symbol is used to denote the averaged parts and the tilded symbol denotes the fluctuating parts. As an example, we show the splitting of the density field (\ref{rho_averaging}), similar expressions have to be written for concentration, pressure and chemical potential.

The equations for the long-term evolution will be obtained by averaging the equations. To accomplish averaging, the following general equalities are required to be used,
\begin{equation}
\overline{\tilde{V}}=0, \quad \overline{\frac{\partial}{\partial t_{-2}}(\dots)}=0,
\end{equation}
where $V$ stands for any quantity.  

Averaging gives the following equations for the processes on diffusive and convective time-scales,
\begin{eqnarray}
\bar{p}_0=0,\quad \bar{\mu}_1=0;\\ 
\nabla \bar{p}_2-Ga_2\boldsymbol{\gamma}=0,\\
\frac{\partial \boldsymbol{u}_2}{\partial t_2}+(\boldsymbol{u}_2\cdot \nabla)\boldsymbol{u}_2+\overline{(\boldsymbol{w}_2\cdot \nabla)\boldsymbol{w}_2}= -\nabla\bar{p}_4 \nonumber \\ +\frac{1}{Re_{-2}}\nabla^2 \boldsymbol{u}_2-Ca_2\nabla C_1\otimes\nabla C_1 - Ga_2 \rho_2 \boldsymbol{\gamma}, \\
\nabla\cdot\boldsymbol{u}_2=0, \label{solenoid} \\
\frac{\partial C_1}{\partial t_2}+(\boldsymbol{u}_2\cdot\nabla)C_1=\frac{1}{Pe_0}\nabla^2\bar{\mu}_3,\\ 
\bar{p}_2=\frac{1}{\phi_1}(-\bar{\mu}_3+2A_2C_1+4C^3_1-Ca_2\nabla^2 C_1),\\
\rho_2=\phi_1 C_1.
\end{eqnarray}
Here, we took into account that $C_1$ and $\rho_2$ are independent of the quick time (\ref{rho2C1}) so the over-bars were omitted for these variables.

Subtracting the averaged parts from the full equations, we obtain the equations for the processes on the quick time-scale,
\begin{eqnarray}
\frac{\partial \tilde{\rho}_4}{\partial t_{-2}}=-\nabla\cdot\boldsymbol{w}_2,\quad \tilde{\rho}_4=\phi_1\tilde{C}_3, \\
\frac{\partial \tilde{C}_3}{\partial t_{-2}}=\frac{1}{Pe_0}\nabla^2\tilde{\mu}_1, \\
\frac{\partial \boldsymbol{w}_2}{\partial t_{-2}}=-\nabla \tilde{p}_0,\quad \tilde{p}_0=\frac{1}{\phi_1}(-\tilde{\mu}_1).
\end{eqnarray}
These equations describe a rapidly decaying process. Re-writing these equations in terms of one variable (e.g. pressure), we find
\begin{equation}
\nabla^2 \tilde{p}_0=\nabla^2 \tilde{p}_{0,in}\exp\left(-\frac{Pe_0}{\phi^2_1}t_{-2}\right),
\end{equation}
where index `$in$' denotes the initial value. If the averaged equations are characterised by the divergence-free fluid velocity (\ref{solenoid}), for the quick processes, the quasi-compressibility effects are essential. But the divergence of the velocity field also rapidly decreases following a similar exponential decay,
\begin{equation}
(\nabla\cdot\boldsymbol{w}_2)=(\nabla\cdot\boldsymbol{w}_{2,in})\exp\left(-\frac{Pe_0}{\phi^2_1}t_{-2}\right).
\end{equation} 

Finally, we note that, firstly, the quick time-scale processes are damping and, secondly, there is no energy injection into fast processes as it could, for example, happen in the case of imposed high-frequency vibrations \cite[][]{Gershuni98, Lyubimov06}. These two facts allow us to draw a conclusion that the short-term processes, even if existed at the initial moment, are to be rapidly damped out and are not to affect the long-term evolution on the convective and diffusive time-scales. 

Omitting indexes and bars, the governing equations for the processes on the convective time scale can be finally written as follows
\begin{equation}\label{NS_final}
\frac{\partial \boldsymbol{u}}{\partial t}+(\boldsymbol{u}\cdot\nabla)\boldsymbol{u} = -\nabla \Pi +\frac{1}{Re}\nabla^2\boldsymbol{u} - C\nabla\mu - Ga \phi C\boldsymbol{\gamma},
\end{equation}
\begin{equation}\label{diff_final}
\frac{\partial C}{\partial t}+(\boldsymbol{u}\cdot\nabla)C=\frac{1}{Pe}\nabla^2 \mu,
\end{equation}
\begin{equation}\label{cont_final}
\nabla\cdot\boldsymbol{u}=0,
\end{equation}
\begin{equation}\label{mu}
\mu\equiv2AC+4C^3-Ca\nabla^2 C.
\end{equation}
Here variable $\Pi$ stands for the modified pressure, value of which can be determined using an incompressibility constraint. 

The derived equations must be supplemented with the following boundary conditions:
\begin{equation}\label{bc}
\boldsymbol{v}=0,\quad \boldsymbol{n}\cdot\nabla C=0,\quad \boldsymbol{n}\cdot\nabla \mu =0.
\end{equation}

Finally, in order to show that no important effects have been lost during our derivation, let us write down the new expression for the total fluid energy:
\begin{widetext}
\begin{equation}\label{energy}
E=\int_V{\left[\frac{\boldsymbol{u}^2}{2}+f_0+\frac{Ca}{2}(\nabla C)^2+ Ga \phi C(\boldsymbol{\gamma}\cdot\boldsymbol{r})\right]dV},\quad f_0=AC^2+C^4.
\end{equation}
\end{widetext}

The finally derived equations (\ref{NS_final}), (\ref{diff_final}) and (\ref{cont_final}) turn out to coincide with the model first used by \cite{Jacqmin99, Jacqmin00} as a computational tool for tracking of the complex transformations of an interfacial boundary between two immiscible liquids. Expression (\ref{Jacq_f0}) was used to define the free energy function. The computational model, adopted by \cite{Jacqmin99,Jacqmin00}, was earlier proposed  by several researchers (for derivation and further references see Refs. \onlinecite{Jasnow96,Anderson98}). Here, we strictly showed that the equations of Jacqmin do, in fact, represent the Boussinesq approximation of the Cahn-Hilliard-Navier-Stokes equations. That is, these equations can be also used as the hydrodynamic model for slow dissolution processes in miscible systems.

\section{Conclusions}

In the current paper, firstly, it was shown that the evolution of a multiphase system of two incompressible slowly miscible liquids is characterised by three typical time scales, namely, the convective and diffusion times and the typical expansion time (\ref{time_scales}). By using the multiple scale method, we filtered out the short-term expansion process and, as a result, the equations for the slow evolution of a binary system were derived. Even if the quick processes are not supported and should not affect the real physical system, the numerical integration of the full Cahn-Hilliard-Navier-Stokes equations would require time-resolution of all processes (the quick processes will be constantly excited due to discretization errors of a numerical scheme) and this would make calculations of long-term dissolution dynamics unfeasible.        

The obtained equations (\ref{NS_final}), (\ref{diff_final}) and (\ref{cont_final}) form a closed mathematical model for a general thermo- and hydro-dynamic evolution of a multiphase system. We should underline, that an applicability of the final model is to be considered in a broader extent compared to what is defined by assumptions (\ref{ratios_conv}), and not, e.g., only for the binary systems in the vicinity of the critical point. For illustration, we may refer the classical Boussinesq equations for thermal convection: despite the fact that such equations are derived in assumption of finite Rayleigh numbers, most interesting convection problems are considered for large Rayleigh numbers ($>1000$).

We should also note that Refs. \onlinecite{Jacqmin99,Jacqmin00,Lowengrub98,Joseph96} are all well known within the scientific community interested in hydrodynamics of multiphase flows. However, these papers contain different sets of governing equations making other researchers to provide their additional justifications for the actual models chosen, see e.g. Refs. \onlinecite{Chen02, Ding07a}. Most of the numerical studies, based on the phase-field approach, use the Jacqmin's model but, in such papers, the evolution of immiscible systems is targeted. For miscible systems, the most popular approach is the set of impurity-like equations with an addition of the Korteweg stress tensor \cite[][]{Chen02, Bessonov04, Chen07}. The main achievement of our work is an establishment of the relation between the full Cahn-Hilliard-Navier-Stokes equations of Lowengrub and Truskinovsky and the divergence-free equations of Jacqmin. We showed that the Jacqmin's equations define the thermo- and hydrodynamic evolution of both immiscible and miscible systems. An important difference of Jacqmin's model from the impurity-like equations lies in the definition of diffusion flux through the gradient of the chemical potential rather than the concentration gradient. Such an amendment takes into account the surface tension effects in calculations of the dissolution rate. This means that the surface tension effects define not only the morphology of the interfacial boundary (the Korteweg stress in the Navier-Stokes equation) but also the diffusion rate. While the first effect is frequently taken into account, the second one is frequently missed out.  

We also would like to note that the derived equations (\ref{NS_final}), (\ref{diff_final}) and (\ref{cont_final}) are not ready yet for a complete analysis of a particular configuration, as the values of the introduced non-dimensional parameters, $\epsilon$, $a$ and $b$, are unknown. In fact, the value of $a$ can be relatively easily estimated from the phase diagram for a particular system. Determination of $\epsilon$ and $b$ would require a more lengthy investigation. Jacqmin published his equations about ten years ago. Since then, these equations were applied to different problems \cite[][]{Jacqmin99, Jacqmin00, Villanueva06, Ding07a, Ding10}, but they always were used for immiscible systems, for which a fluid behaviour in the limits of $\epsilon\rightarrow0$ and $b\rightarrow\infty$ is sought. Technically, the equations are successively solved for several values of parameters to reproduce the needed limiting behaviour of an immiscible interface, see e.g. Refs. \onlinecite{Jacqmin99, Jacqmin00, Ding07a, Ding10}. For the case of miscible systems, the values of $\epsilon$ and $b$ are finite and should be obtained from a comparison of the numerical solution with the experimental data. In some recent experiments, the estimations and measurements of the surface tension coefficients for several particular miscible binary systems became available \cite{Petitjeans96, Pojman06, Zoltowski07}. These experiments also contain detailed description of the dissolution dynamics. Such information (time evolution of the surface tension coefficient and of the dissolution rate) should be sufficient for obtaining the missing values of $\epsilon$ and $b$. This is the aim of the author's current research work. 

\begin{acknowledgments}
This work is done within the framework of the EPSRC project EP/G014337 `Pore-Level Network Modelling of the Miscible Displacement.' The author wishes to thank Prof. D.V. Lyubimov from the Perm State University for many fruitful discussions giving many ideas of this work.
\end{acknowledgments}

\bibliography{Boussinesq}

\end{document}